\documentclass[aps, pre, twocolumn, superscriptaddress]{revtex4-1}
\usepackage{bm}
\usepackage{amsmath}
\usepackage{amsfonts}
\usepackage[utf8]{inputenc}

\begin{document}

\title{Emergence of Tsallis statistics as a consequence of invariance}

\author{Sergio Davis}
\email{sergio.davis@cchen.cl}
\affiliation{Comisión Chilena de Energía Nuclear, Casilla 188-D, Santiago, Chile}
\affiliation{Departamento de F\'isica, Facultad de Ciencias Exactas, Universidad Andres Bello. Sazi\'e 2212, piso 7, 8370136, Santiago, Chile.}

\author{Gonzalo Guti\'errez}
\affiliation{Grupo de Nanomateriales, Departamento de F\'{i}sica, Facultad de Ciencias, Universidad de Chile, Casilla 653, Santiago, Chile}

\date{\today}

\begin{abstract}
For non-equilibrium systems in a steady state we present two necessary and sufficient conditions for the emergence of
$q$-canonical ensembles, also known as Tsallis statistics. These conditions are invariance requirements over the definition
of subsystem and environment, and over the joint rescaling of temperature and energy. Our approach is complementary to the notions
of Tsallis non-extensive statistics and Superstatistics.
\end{abstract}



\maketitle

\section{Introduction}

One of the most ubiquitous statistical distributions for non-equilibrium, steady-state systems is known as the $q$-canonical ensemble.
Systems described by $q$-canonical distributions are common in Nature, as non-equilibrium steady states in plasmas~\cite{Lima2000,Livadiotis2015}, fluids under
turbulence~\cite{Jung2004}, astrophysical systems where gravitational interactions are dominant~\cite{Du2004b}, high-energy collisions~\cite{Cleymans2013},
among several others. The $q$-canonical ensemble is a statistical model described by two parameters, $\beta_0$ and $q$, which assigns a probability density to the
microstates $\bm{\Gamma}$ given by

\begin{equation}
P(\bm{\Gamma} | \beta_0, q) = \frac{1}{Z_q(\beta_0, q)}\left[1-(1-q)\beta_0 H(\bm{\Gamma})\right]_{+}^{\frac{1}{1-q}},
\label{eq_qcanon_microstates}
\end{equation}
where $\left[x\right]_+ = \max(x, 0)$, in such a way that it converges to the canonical ensemble in the limit $q \rightarrow 1$, that is

\begin{equation}
P(\bm{\Gamma} | \beta_0, q = 1) = \frac{1}{Z(\beta_0)}\exp\left(-\beta_0 H(\bm{\Gamma})\right).
\label{eq_canon}
\end{equation}

The central problem in terms of fundamental statistical physics is to explain the origin of this family of non-canonical distributions. In 1988
Constantino Tsallis~\cite{Tsallis1988, Tsallis2009} proposed a generalization of Boltzmann-Gibbs statistical mechanics now known as \emph{non-extensive statistical mechanics},
in which instead of the Gibbs-Shannon entropy one maximizes the generalized entropy

\begin{equation}
S_q = \frac{1}{q-1}\left(1-\int d\bm{\Gamma}\; p(\bm{\Gamma})\right)^q
\end{equation}
subjected to constraints. After Tsallis' work, $q$ is known as the non-extensivity parameter or entropic index, and the $q$-canonical statistics as Tsallis statistics.
More recently, alternative mechanisms which explain the emergence of $q$-canonical ensembles have been proposed, most prominently the idea of Superstatistics~\cite{Beck2003, Beck2004, Sattin2006}.
The superstatistical framework considers a system having a statistical distribution of temperatures described by the probability density $P(\beta|\mathcal{S})$, so that the microstates
are weighted with the probability distribution

\begin{equation}
P(\bm{\Gamma}|\mathcal{S}) = \int_0^\infty d\beta P(\beta|\mathcal{S})\left[\frac{\exp(-\beta H(\bm{\Gamma}))}{Z(\beta)}\right].
\label{eq_superstat}
\end{equation}

In the particular case when there is a single temperature, i.e. $P(\beta|\mathcal{S})$ is a Dirac delta function, we recover the canonical ensemble,
whereas the $q$-canonical ensemble arises when the factor $f(\beta)=P(\beta|\mathcal{S})/Z(\beta)$ is given by a Gamma distribution~\cite{Beck2003}.

In this work we propose a third mechanism, an alternative to the maximization of Tsallis entropy and Superstatistics. We show that the $q$-canonical ensemble arises naturally
from invariance considerations, when expressed using the concept of the \emph{fundamental temperature function}.

This work is organized as follows. After this introduction, in Section \ref{sect_notation} we give a brief description of generalized, steady-state ensembles and introduce the concept of fundamental
temperature. In Section \ref{sect_postulates} we state the postulates leading to the $q$-canonical ensemble, while Section \ref{sect_proof} provides the detailed proof of our
main result. We close with some concluding remarks in Section \ref{sect_concluding}.

\section{The generalized Boltzmann factor and the fundamental temperature}
\label{sect_notation}

For the canonical ensemble (Eq. \ref{eq_canon}) the probability of having an energy $E$ is given by

\begin{align}
P(E|\beta) & = \Big<\delta(E-H(\bm{\Gamma}))\Big>_\beta \nonumber \\
           & = \int d\bm{\Gamma}\delta(E-H(\bm{\Gamma}))\left[\frac{\exp(-\beta H(\bm{\Gamma}))}{Z(\beta)}\right] \nonumber \\
           & = \frac{\exp(-\beta E)}{Z(\beta)}\Omega(E),
\label{eq_energy_canon}
\end{align}
where $\Omega(E)$ is the density of states,

\begin{equation}
\Omega(E) = \int d\bm{\Gamma}\delta(E-H(\bm{\Gamma})).
\end{equation}

In this work we will be focusing on more general \emph{steady states} $P(\bm{\Gamma}|\mathcal{S})$ such that

\begin{equation}
P(\bm{\Gamma}|\mathcal{S})=\rho(H(\bm{\Gamma})).
\label{eq_SS}
\end{equation}
where $\rho(E)$ will be referred to as the generalized Boltzmann factor. The probability density for the
energy of the system in such an ensemble is given by

\begin{equation}
P(E|\mathcal{S}) = \rho(E)\Omega(E).
\label{eq_energy_rho}
\end{equation}

Simple inspection shows that superstatistical ensembles, given by Eq. \ref{eq_superstat}, are particular cases of this form,
where the generalized Boltzmann factor is

\begin{equation}
\rho(E) = \int_0^\infty d\beta \left[\frac{P(\beta|\rho)}{Z(\beta)}\right]\exp(-\beta E),
\end{equation}
which is the Laplace transform of the function $f(\beta)=P(\beta|\rho)/Z(\beta)$.

For a steady state described by Eq. \ref{eq_SS}, let us define the fundamental inverse temperature
function~\cite{Loguercio2016, Palma2016, Davis2018}, denoted by $\beta_F(E)$, as the derivative

\begin{equation}
\beta_F(E) = -\frac{\partial}{\partial E}\ln \rho(E).
\label{eq_def_fund}
\end{equation}

Although this quantity has appeared before in the literature as an effective inverse temperature induced by the surroundings
of a system~\cite{Velazquez2009,Velazquez2009a}, in this work it will take a broader and more important meaning, as a
generalization of the inverse temperature for any system in a steady state in the sense of Eq. \ref{eq_SS}.

Knowledge of $\beta_F(E)$ is completely equivalent to knowledge of $\rho(E)$, as we can recover it by simple integration
and normalization, and thus it also completely describes the ensemble. However, it has usually a simpler form than
$\rho$ and can be read in a more intuitive way. In the context of superstatistics, $\beta_F(E)$ has a clear interpretation:
it is the conditional expectation of the superstatistical parameter $\beta$ at a given energy $E$, that is,

\begin{equation}
\beta_F(E) = \big<\beta\big>_{E, \mathcal{S}}.
\label{eq_fund_interp}
\end{equation}

The proof of this equivalence is given in the Appendix.

As we have defined it, $\beta_F(E)$ is an ensemble-dependent function of the energy of the system. The only case where this
function is a constant is of course the canonical ensemble, which can be seen clearly from Eq. \ref{eq_fund_interp}, as in
this case there is a single superstatistical (inverse) temperature, namely $\beta_0$, and the expectation
$\big<\beta\big>_E=\beta_0$, independent of the energy. Also by integration of Eq. \ref{eq_def_fund}, $\beta_F(E)=\beta_0$
implies $\rho(E) \propto \exp(-\beta_0 E)$.

A more interesting case to consider is the $q$-canonical ensemble, given by Eq. \ref{eq_qcanon_microstates}. Now the generalized Boltzmann factor
$\rho(E)$ is

\begin{equation}
\rho(E) = \frac{1}{Z_q(\beta_0)}\left[1-(1-q)\beta_0 E\right]_{+}^{\frac{1}{1-q}} = \rho(E; \beta_0, q)
\label{eq_qcanon}
\end{equation}
and the fundamental inverse temperature is given by~\cite{Davis2013}

\begin{equation}
\beta_F(E; \beta_0, q) = \frac{\beta_0}{1-(1-q)\beta_0 E}.
\label{eq_qcanon_betaf}
\end{equation}

In terms of $\beta_F(E)$ it is straightforward to take the limit $q=1$, in which $\beta_F(E) \rightarrow \beta_0$ (i.e. we recover the canonical ensemble).

Some of the functional forms for $\beta_F$ for different ensembles are given in Table \ref{tbl_models}.

\begin{table}
\begin{tabular}{|c|c|c|}
\hline
Ensemble & $\rho(E)$ & $\beta_F(E)$ \\
\hline
Canonical & $\exp(-\beta_0 E)/Z(\beta_0)$ & $\beta_0$ \\
\hline
$q$-canonical & $\frac{1}{\eta(\beta_0, q)}\left[1-(1-q)\beta_0 E\right]_+^{\frac{1}{1-q}}$ & $\beta_0/(1-(1-q)\beta_0 E)$ \\
\hline
Gaussian & $\frac{1}{Z(a, E_t)}\exp(-a(E-E_t)^2)$ & $2a(E-E_t)$ \\
\hline
\end{tabular}
\caption{Generalized Boltzmann factor $\rho(E)$ and fundamental inverse temperature function $\beta_F(E)$ for the
canonical, $q$-canonical and Gaussian~\cite{Challa1988a,Johal2003} ensembles.}
\label{tbl_models}
\end{table}

\section{Statement of the postulates}
\label{sect_postulates}

Our aim in this section is to describe the $q$-canonical ensemble with a minimal set of postulates, based on our
definition of the fundamental temperature. In fact, we will show that only two such postulates suffice to recover the
$q$-canonical form of $\beta_F$ given in Eq. \ref{eq_qcanon_betaf}.

The first postulate asserts that the same fundamental temperature function applies to the whole of a system, as well as its parts.
More precisely, for a system with energy $E$ which can be divided into two contributions, namely $E=E_1+E_2$, we impose that

\begin{equation}
\beta_F(E_1+E_2; \beta_0) = \beta_F(E_1; \beta_F(E_2; \beta_0)).
\label{eq_first_postulate}
\end{equation}

A simple interpretation of this is that part of the system, encoded in $E_2$, is ``absorbed'' into the environment that
affects the rest (represented by $E_1$), and that environment is no longer constant, but depends on the fluctuations of
$E_2$ \emph{through the same function} $\beta_F$.

Setting $E_1=E_2=0$ in Eq. \ref{eq_first_postulate} leads to the boundary condition $\beta_F(0; \beta_0) = \beta_0$ for any value of $\beta_0$,
that is, the limit of low energy always corresponds to the canonical ensemble.



The second postulate requires the existence of a function $D(x)$ such that

\begin{equation}
\beta_F(E; \beta_0) E = D(\beta_0 E).
\label{eq_second_postulate}
\end{equation}

In this way, the physical properties of the system only depend on the product $\beta E$ (the ratio \emph{energy/temperature}).
Rescaling the energy and the temperature simultaneously by the same factor $\alpha$ cannot have any effect on the description of the system.
This can be easily seen in Eq. \ref{eq_cvt_rho}, where rescaling $E \rightarrow \alpha E$ and $\beta_F \rightarrow \beta_F/\alpha$ will not
change the expectation of $\big<\nabla\cdot \bm{v}\big>_\rho$.

It is a simple exercise to check that, for the $q$-canonical fundamental temperature given in Eq. \ref{eq_qcanon_betaf} both
requirements (Eqs. \ref{eq_first_postulate} and \ref{eq_second_postulate}) hold for any value of $q$. We will go further, and in the
next section we will show that the $q$-canonical ensemble is the only possible model compatible with both postulates simultaneously.

\section{Proof of the uniqueness of the $q$-canonical form}
\label{sect_proof}

We start by assuming the first postulate,

\begin{equation}
\beta_F(h+g; \beta_0) = \beta_F(h; \beta_F(g; \beta_0)),
\end{equation}
and recognizing its validity for any combination of values of $h$, $g$ and $\beta_0$. By differentiation with respect to
$h$ and $g$, we find a system of differential equations, namely

\begin{eqnarray}
\frac{\partial \beta_F(A; \beta_0)}{\partial A}\Big|_{A=h+g}\!\!\!\!\!\!\!\!\!\!\!\! = \frac{\partial \beta_F(h; \beta_F(g; \beta_0))}{\partial h}, \label{eq_system_1} \\
\frac{\partial \beta_F(A; \beta_0)}{\partial A}\Big|_{A=h+g}\!\!\!\!\!\!\!\!\!\!\!\! = \frac{\partial \beta_F(h; B)}{\partial
B}\Big|_{B=\beta_F(g;\beta_0)}\!\!\!\!\!\!\frac{\partial \beta_F(g; \beta_0)}{\partial g}.
\label{eq_system_2}
\end{eqnarray}

Combining Eq. \ref{eq_system_1} and \ref{eq_system_2} we get

\begin{equation}
\frac{\partial \beta_F(g; \beta_0)}{\partial g} = \frac{\left(\frac{\partial \beta_F(h; B)}{\partial h}\right)}{\left(\frac{\partial \beta_F(h; B)}{\partial B}\right)}\Big|_{B=\beta_F(g;\beta_0)},
\end{equation}
which implies, on the one hand, that the right-hand side is a function of $\beta_F(g;\beta_0)$, that is,

\begin{equation}
\frac{\partial \beta_F(g; \beta_0)}{\partial g} = F(\beta_F(g; \beta_0)),
\label{eq_partial_1}
\end{equation}
and also that it is independent of $h$, therefore

\begin{equation}
\frac{\partial \beta_F(h; B)}{\partial h} = F(B)\left(\frac{\partial \beta_F(h; B)}{\partial B}\right).
\label{eq_partial_3}
\end{equation}

Combining Eq. \ref{eq_partial_1} with $g=h$ and Eq. \ref{eq_partial_3}, we obtain

\begin{equation}
\frac{\partial \beta_F(h; \beta_0)}{\partial \beta_0} = \frac{F(\beta_F(h; \beta_0))}{F(\beta_0)}.
\label{eq_partial_2}
\end{equation}

Now we incorporate the second postulate, $\beta_F(h; \beta_0) h =D(\beta_0 h)$ and require the equality of the
second-order cross derivatives,

\begin{equation}
\frac{\partial^2 \beta_F}{\partial h\partial \beta_0} = F'\left(\frac{D(\beta_0 h)}{h}\right)D'(\beta_0 h)
\end{equation}
and

\begin{equation}
\frac{\partial^2 \beta_F}{\partial h\partial \beta_0} = \frac{F'\left(\frac{D(\beta_0 h)}{h}\right)}{F(\beta_0)}
\left[\frac{\beta_0 D'(\beta_0 h)}{h} - \frac{D(\beta_0 h)}{h^2}\right],
\end{equation}
leading to

\begin{equation}
\frac{F(\beta_0)}{\beta_0^2} = \frac{1}{\beta_0 h} - \frac{D(\beta_0 h)}{D'(\beta_0 h)(\beta_0 h)^2}.
\end{equation}

As the left-hand side does not depend on $h$, it follows that both sides are equal to a constant, $K$. Then, the
function $F$ is given by $F(x)=Kx^2$ and also

\begin{equation}
D'(x) = \frac{D(x)}{x(1-Kx)},
\end{equation}
with general solution,

\begin{equation}
D(x) = \frac{\lambda x}{1-Kx}.
\end{equation}
where $\lambda$ is an integration constant, to be determined. We now have, for the fundamental inverse temperature,

\begin{equation}
\beta_F(h; \beta_0) = \frac{D(\beta_0 h)}{h} = \frac{\lambda \beta_0}{1-K \beta_0 h}.
\end{equation}

As $\beta_F(0; \beta_0)=\beta_0$, the integration constant $\lambda$ must be equal to one, and therefore the only fundamental inverse
temperature compatible with Eqs. \ref{eq_first_postulate} and \ref{eq_second_postulate} has the $q$-canonical form

\begin{equation}
\beta_F(h; \beta_0) = \frac{\beta_0}{1-(1-q)\beta_0 h},
\end{equation}
after the identification of $K$ with $1-q$. We can check as well that the function $F(x)=Kx^2$ leads to the correct
partial derivatives in Eqs. \ref{eq_partial_1} and \ref{eq_partial_2} for the $q$-canonical,

\begin{align}
\frac{\partial \beta_F(h; \beta_0)}{\partial h} & = (1-q)\beta_F(h; \beta_0)^2 = F(\beta_F(h; \beta_0)), \nonumber \\
\frac{\partial \beta_F(h; \beta_0)}{\partial \beta_0} & = \frac{\beta_F(h; \beta_0)^2}{\beta_0^2} = \frac{F(\beta_F(h; \beta_0))}{F(\beta_0)}.
\end{align}

\section{Concluding remarks}
\label{sect_concluding}

We have shown that the first and second postulates, Eqs. \ref{eq_first_postulate} and \ref{eq_second_postulate},
are necessary and sufficient conditions for the emergence of $q$-canonical probability distributions. These postulates
are in fact invariance requirements, on the partition between a subsystem and its environment, and under joint rescaling
of energy and temperature. Note that, unlike Tsallis statistics or Superstatistics, we have not assumed \emph{a priori} the
existence of a parameter $q$ in our postulates; it rather appears naturally as a parametrization of the family of joint solutions
of Eq. \ref{eq_first_postulate} and \ref{eq_second_postulate}.

It is interesting to note that the first postulate imposes that the same fundamental temperature function, even the same
$q$ index, has to be used for subsystems in contact, and for the whole. This is reminiscent of the issue raised by
Nauenberg~\cite{Nauenberg2003} about equilibrium between systems with different values of $q$.

\section*{Acknowledgements}

SD and GG gratefully acknowledge funding from FONDECYT grant 1171127. SD also acknowledges funding from Anillo ACT-172101 grant.

\appendix
\section{Properties of the fundamental inverse temperature function}

A model with fundamental inverse temperature $\beta_F$ has a generalized equipartition theorem given by~\cite{Palma2016}

\begin{equation}
\Big<\nabla\cdot \bm{v}\Big>_\mathcal{S} = \Big<\beta_F(H)\bm{v}\cdot \nabla H\Big>_\mathcal{S},
\label{eq_cvt_rho}
\end{equation}
whose equivalent, in the superstatistical formalism, is the identity

\begin{equation}
\big<\nabla\cdot \bm{v}\big>_\mathcal{S} = \int_0^\infty d\beta P(\beta|\rho)\cdot \beta\big<\bm{v}\cdot\nabla H\big>_\beta
  = \big<\beta \bm{v}\cdot \nabla H\big>_\mathcal{S},
\label{eq_cvt_super}
\end{equation}
constructed from taking expectation of the canonical equipartition theorem~\cite{Davis2012}, $\big<\nabla \cdot \bm{v}\big>_\beta =
\beta\big<\bm{v}\cdot\nabla H\big>_\beta$ under $P(\beta|\mathcal{S})$. As Eqs. \ref{eq_cvt_rho} and \ref{eq_cvt_super} are both valid for
any field $\bm{v}$, it follows that

\begin{equation}
\Big<\beta_F(H)\bm{v}\cdot\nabla H\Big>_\mathcal{S} = \Big<\beta\bm{v}\cdot \nabla H\Big>_\mathcal{S},
\end{equation}
for any $\bm{v}=\bm{v}(\bm{\Gamma})$. Using the particular choice

\begin{equation}
\bm{v}=\frac{g(\bm{\Gamma})\bm{\omega}}{\bm{\omega}\cdot\nabla H},
\end{equation}
we see that

\begin{equation}
\Big<\beta_F(H)g(\bm{\Gamma})\Big>_\mathcal{S} = \Big<\beta\;g(\bm{\Gamma})\Big>_\mathcal{S},
\end{equation}
for any function $g(\bm{\Gamma})$. In particular, for $g(\bm{\Gamma})=\delta(E-H(\bm{\Gamma}))$ we obtain that $$\beta_F(E) = \big<\beta\big>_{E, \mathcal{S}}.$$

As a direct proof of the last assertion, consider the energy distribution functions for the canonical (Eq. \ref{eq_energy_canon}) and the ensemble described
by $\rho(E)$ in Eq. \ref{eq_energy_rho}, and then use Bayes' theorem to construct

\begin{equation}
P(\beta|E, \rho) = \frac{P(\beta|\rho)P(E|\beta)}{P(E|\rho)} = \frac{P(\beta|\rho)\exp(-\beta E)}{\rho(E)Z(\beta)}.
\end{equation}

Taking the expectation of $\beta$ with this probability distribution yields,

\begin{align}
\big<\beta\big>_{E,\rho} & = \int_0^\infty d\beta P(\beta|E,\rho)\cdot \beta \nonumber \\
& = \int_0^\infty d\beta \left[\frac{P(\beta|\rho)\exp(-\beta E)}{\rho(E)Z(\beta)}\right]\cdot \beta \nonumber \\
& = -\frac{1}{\rho(E)}\int_0^\infty d\beta \frac{P(\beta|\rho)}{Z(\beta)}\frac{\partial}{\partial E}\left(\exp(-\beta E)\right) \nonumber \\
& = -\frac{1}{\rho(E)}\frac{\partial \rho(E)}{\partial E} = \beta_F(E).
\end{align}



\end{document}